\documentclass[sigconf,nonacm]{acmart}
\usepackage{array}
\usepackage{graphicx}
\usepackage{enumitem}
\usepackage{tikz}
\usepackage[most]{tcolorbox}

\AtBeginDocument{%
  }

\begin{document}

\title{Toward Agentic Software Project Management: A Vision and Roadmap}

\author{Lakshana Iruni Assalaarachchi*}
\email{lakshana.assalaarachchi@monash.edu}
\orcid{https://orcid.org/0009-0006-1848-3233}
\affiliation{%
  \institution{Monash University}
  \city{Clayton}
  \state{Victoria}
  \country{Australia}
}
\author{Zainab Masood}
\email{zmasood@psu.edu.sa}
\orcid{https://orcid.org/0000-0002-3592-8250}
\affiliation{%
  \institution{Prince Sultan University}
  \city{Riyadh}
  \country{Saudi Arabia}
}
\author{Rashina Hoda}
\email{rashina.hoda@monash.edu}
\orcid{https://orcid.org/0000-0001-5147-8096}
\affiliation{%
  \institution{Monash University}
  \city{Clayton}
  \state{Victoria}
  \country{Australia}
}
\author{John Grundy}
\email{john.grundy@monash.edu}
\orcid{https://orcid.org/0000-0003-4928-7076}
\affiliation{%
  \institution{Monash University}
  \city{Clayton}
  \state{Victoria}
  \country{Australia}
}

\renewcommand{\shortauthors}{Assalaarachchi et al.}

\begin{abstract}
  With the advent of agentic AI, Software Engineering is transforming to a new era dubbed Software Engineering 3.0. Software project management (SPM) must also evolve with such transformations to boost successful project completion, while keeping humans at the heart of it. Building on our preliminary ideas of \textit{"agentic SPM"}, and supporting literature, we present our vision of an \textit{"Agentic Project Manager (PM)"} as a multi-agent system for SPM 3.0. They will work like a \textit{“junior project manager”}, or an \textit{“intern project manager”} collaboratively with software teams. We introduce four working modes, with varying autonomy levels to choose from, based on the SPM task. This addresses concerns with ethics, accountability, and trust related to agentic PMs. We also share insights on human PM role evolution and new skill requirements as a \textit{“strategic leader”} and a \textit{“coach”} for humans and agents. While creating the foundation for agentic SPM research, we present a research agenda for the wider research community.
\end{abstract}




\maketitle
\section{Introduction}
Generative artificial intelligence (GenAI) advancements are leading us towards an "Agentic AI" era \cite{Mori2025}. Agentic AI refers to autonomous AI systems that achieve complex goals through autonomous decision making, proactive task execution with minimum human involvement, learn, and adapt by working with humans and systems. This makes agentic AI a \textit{“living system”} in contrast to other AI tools \cite{White2024}.

\renewcommand{\thefootnote}{}
\footnote{This version is the author's preprint of the article accepted for AGENT workshop at ICSE 2026}
\addtocounter{footnote}{-1}

Recent studies envision Software Engineering 3.0 (SE 3.0, also dubbed Agentic SE), a new era of SE, with agentic AI as teammates working collaboratively with humans in the software development \cite{Li2025}. Studies on the use of agentic SE have started identifying potential to assist in software development activities such as coding and testing \cite{Abrahao2025, Li2025}. These findings complement the ongoing practitioner discussions about the use of AI agents in SE, such as Zapier\footnote{https://zapier.com/ai}, for task automation \cite{Mori2025}, AI-powered assistants such as GitLab Duo\footnote{https://about.gitlab.com/gitlab-duo/}, Anthropic’s Claude Code\footnote{https://www.claude.com/product/claude-code}, Sourcegraph Amp\footnote{https://sourcegraph.com/amp} \cite{Karatas2025}, and Rovo Dev\footnote{https://www.atlassian.com/software/rovo-dev} in code generation and software testing. With this paradigm shift, there is a need to redefine existing methodologies, roles, practices, and artifacts across the software development life cycle (SDLC), without limiting it to programming \cite{hoda2025agenticsoftwareengineeringcode}.

We also noted that practitioners anticipate agentic AI is coming closer to becoming a reality in software project management (SPM). A recent report on AI in Project Management by Project Management Institute (PMI Sweden chapter) \cite{Nilsson2025} has identified Information Technology (IT) as one of the top three sectors that uses AI in project management. As part of the ongoing AI evolution, it is expected to have virtual project assistants to facilitate human project manager (PM) \cite{Nilsson2025}. Some practitioners presume a role like \textit{“AI-enhanced project manager”} by 2030  \cite{Masood2025}.

We concluded our recent review of practitioner literature on the use of GenAI in SPM with a brief vision of agentic SPM, where AI agents act as assistants to human PM in different SPM tasks, and the human PM interacts not just with people and processes but also with those SPM agents \cite{Assalaarachchi2025a}. However, with evolving agentic AI concepts, this raises concerns about coordination overheads and the need for integrated or unified agentic systems \cite{electronics14010087, hoda2025agenticsoftwareengineeringcode}. Some sources debate that simply adopting agentic AI as a trend can cause project failures due to high cost, lack of business value mapping, low acceptance by users, or poor risk management \cite{Gartner2025}. Lack of focus on such aspects has also been reported to cause failure of GenAI investments in more than 95\% of organizations \cite{MIT2025}. Therefore, we address the following research questions in this vision paper. 

\begin{enumerate} [leftmargin=12pt]
    \item \textit{How SPM will evolve with AI to facilitate agentic SE?}
    \item \textit{How should an ethical multi-agent project management assistant be designed and used while balancing autonomy?}
    \item \textit{How should the human PM role evolve to facilitate responsible human-agent interactions?}
\end{enumerate}

Referring to emerging literature, we present the \textbf{\textit{"SPM with AI Roadmap"}} (in section \ref{SPMRoadmap section}), and build upon our earlier vision to propose an ethical \textbf{\textit{"agentic PM"}} along with four \textbf{\textit{"working modes"}} that adjust the autonomy of agentic PM based on each SPM task. We elaborate on how and why human oversight should be provided at each working mode to promote responsible human-agent interactions (in section \ref{agenticPM}). We also provide insights on human PM role evolution with skills requirements in the agentic SPM era (in section \ref{humanPMRole}). Our proposed vision will guide AI developers in developing a \textit{trustworthy} agentic PM to increase the level of acceptance and adoption. Current (human) PMs can gain an understanding of their role evolution and upskilling requirements needed to facilitate better human-agent collaboration in the agentic SPM era. Furthermore, we provide a research agenda based on this vision (in section \ref{researchAgenda}).

\section{Related Works}

\subsection{Agentic AI}
The  pubic launch of ChatGPT in November 2022 made revolutionary changes across multiple industries and sectors \cite{Sapkota2026}. These GenAI tools are now evolving into AI Agents and Agentic AI \cite{Mori2025, Sapkota2026}. 
AI agents are designed as an individual entity that automates a particular tasks with own reasoning capabilities \cite{Hughes2025, Sapkota2026, White2024}. On the other hand, Agentic AI is evolving as a solution for the coordination overhead issues when working with single agents for multiple tasks. Agentic AI is considered a living system given its advanced and unique capabilities in achieving complex goals through autonomous decision making, proactive task execution with minimum human involvement, multi-agent coordination, and learning and adaptability \cite{Sapkota2026, White2024}. Researchers in the SE discipline have started exploring the possibility of agentic SE, a new era in SE supported by agentic teammates, with ongoing agentic AI trends in the software industry \cite{hoda2025agenticsoftwareengineeringcode}.

\subsection{Agentic SE}
A recent vision paper envision Software Engineering 3.0 (SE 3.0, also dubbed Agentic SE), a new era of SE, with agentic AI as teammates working collaboratively with humans in the software development \cite{Hassan2025}. They highlight the need for redefining the SE roles \cite{Hassan2025} and how humans collaborate with agentic teammates to reach  maximum potential \cite{White2024}. Agentic AI is not expected to replace existing roles but will require redefining those roles to have a clear understanding of synergy between human teammates and agentic teammates \cite{Li2025, Hassan2025}, while keeping humans at the heart of the software development life cycle ~\cite{Abrahao2025}. However, most of these current studies focus only on agentic AI in programming and testing \cite{Abrahao2025, Li2025, hoda2025agenticsoftwareengineeringcode}. SPM, must also evolve when restructuring the SDLC with agentic AI, as it facilitates successful project completion, while managing available resources and constraints \cite{Abrahao2025, Assalaarachchi2025a, hoda2025agenticsoftwareengineeringcode}. Recent Roadmap article presents SPM as one of the outer loop activities in SDLC that they expect continuation of human role assisted by AI rather than full automation \cite{Abrahao2025}. Therefore, it is crucial to explore how SPM could evolve to support agentic SE and identify responsible human-agent interactions.

\subsection{Agentic AI and SPM}
Several studies have identified the potential of AI and GenAI tools to assist PMs in the automation of routine tasks such as SPM artifact creation \cite{Dam2019, Kardum2025, Assalaarachchi2025a}, predictive analytics \cite{Dam2019, Assalaarachchi2025a}, data-driven decision making \cite{Dam2019, Kardum2025, Assalaarachchi2025a}, enhancing communication and collaboration \cite{Assalaarachchi2025a}, and better risk management \cite{Dam2019, Assalaarachchi2025a}, supporting PMs in saving more time for strategic activities \cite{Dam2019, Kardum2025, Assalaarachchi2025a}, and achieving project success \cite{Dam2019, Assalaarachchi2025a}. Software practitioners have started to perceive AI as an assistant or a copilot rather than a tool \cite{Assalaarachchi2025a}.   

A vision paper in 2023 (pre-GenAI) has proposed a vision of \textit{augmented agile} with a conceptual \textit{“agile co-pilot”}, a human-centered assistant in agile project management. They emphasize the need for having human-centered values embedded in AI tools and understanding the roles of the human PM and AI assistant rather than using AI as a replacement for the human PM role \cite{Hoda2023}.  Building upon this vision and considering emerging practitioner discussions, we recently proposed the idea of \textbf{\textit{"agentic SPM"}} with evolving AI capabilities. Our review study focused on reporting the use of GenAI in SPM using practitioner literature. As a conclusion, we presented the idea of having AI agents as assistants to the human PM, proactively executing SPM tasks and decision making with some human oversight. We highlighted the need to expand on that vision further to identify practitioners' perceptions towards agentic SPM, and task delegation between human PM and SPM agents to ensure ethical human-AI collaborations \cite{Assalaarachchi2025a}. 

Having SPM agents as \textit{"assistants"} or \textit{"junior PMs"} can support human PM with saving time from routine tasks to focus on strategic tasks \cite{Assalaarachchi2025a, Hughes2025, Masood2025, Gress2025}. It is also expected to facilitate data-driven decision making \cite{Hughes2025, Gress2025}, increases efficiency and productivity \cite{Hughes2025, Assalaarachchi2025a, Gress2025}, and achieve project success \cite{Masood2025, Hughes2025, Gress2025}. In addition to those expected benefits, practitioners also discuss possible challenges with agentic SPM such as privacy and ethics, accountability, trust, skills gap, and fear about job security \cite{Gress2025}. Some sources also highlight the need for developing ethical AI systems and governance frameworks given the increase in risks when increasing AI autonomy \cite{Boinodiris2025, Hughes2025, electronics14010087,SPIEGLER2026103109}. A recent study has demonstrated that losing human control over an AI system could create a negative impact on the society \cite{SPIEGLER2026103109}. Similarly, a recent IBM article also explains that agentic AI could raise more risks and ethical concerns with increasing autonomy. They emphasize the importance of enhancing human accountability and ethical oversight in addition to technological guardrails along with agentic AI evolution \cite{Boinodiris2025}. 

In alignment to these discussions, a recent vision paper proposes autonomy levels for AI agents by identifying autonomy in AI agents as a \textit{“double-edged sword”} that also causes concerns while providing the benefits \cite{Feng2025}. They propose five autonomy levels with user involvement from the minimum level as \textit{“an observer”} to higher user involvement as \textit{“an operator”}. In that study, they propose more autonomy for the agent with just human monitoring under the  \textit{"Level 5: User as an Observer"}. Then they propose \textit{"Level 4: User as an Approver", "Level 3: User as a Consultant", "Level 2: User as a Collaborator",} and \textit{"Level 1: User as an Operator",} which require more user involvement and the lowest autonomy for the agent. Along these levels, from level 5 to level 1, autonomy of the agent decreases while increasing the user involvement. They expect this model to guide future vision of agentic AI design and usage across various sectors. That would help promote responsible human-agent interactions, address concerns on ethics, accountability, trust, and fears of job security \cite{Feng2025}.  

Another recent study on multi-agent systems for scaled agile projects has raised concerns about coordination overhead when single AI agents augment individual tasks \cite{electronics14010087}. Studies recommend future visions to build around interconnected multi-agent systems which are more proactive in task management, suitable for complex goals and dynamic environments, capable of reasoning, learning, and adapting \cite{Sapkota2026, Acharya2025} to facilitate the envisioned agentic SE evolution \cite{Abrahao2025, Hassan2025, hoda2025agenticsoftwareengineeringcode}. Therefore, in this paper, we extend our prior vision to accommodate such a multi-agent system beyond individual agents and introduce the idea of an \textbf{\textit{"agentic project manager (PM)"}}, an adaptable and ethical multi-agent system for SPM that works like an \textit{"intern PM"} or a \textit{"junior PM"}. 

We also note the need for clearer task delegation between human PM and agentic PM, along with defining appropriate levels of autonomy for each SPM task performed by agentic PM \cite{hoda2025agenticsoftwareengineeringcode}. Such an understanding is crucial to promote appropriate trust and acceptance of agentic PMs by humans \cite{Assalaarachchi2025a}. In doing so, we expect the human PM to take the role of a \textit{"coach"} who guides, supervises, and facilitates the team, including agentic teammates, towards successful project completion. 

\section{Agentic SPM} \label{sec3}
\subsection{SPM with AI Roadmap} \label{SPMRoadmap section}

\begin{figure} 
\centerline{\includegraphics[width=20pc]{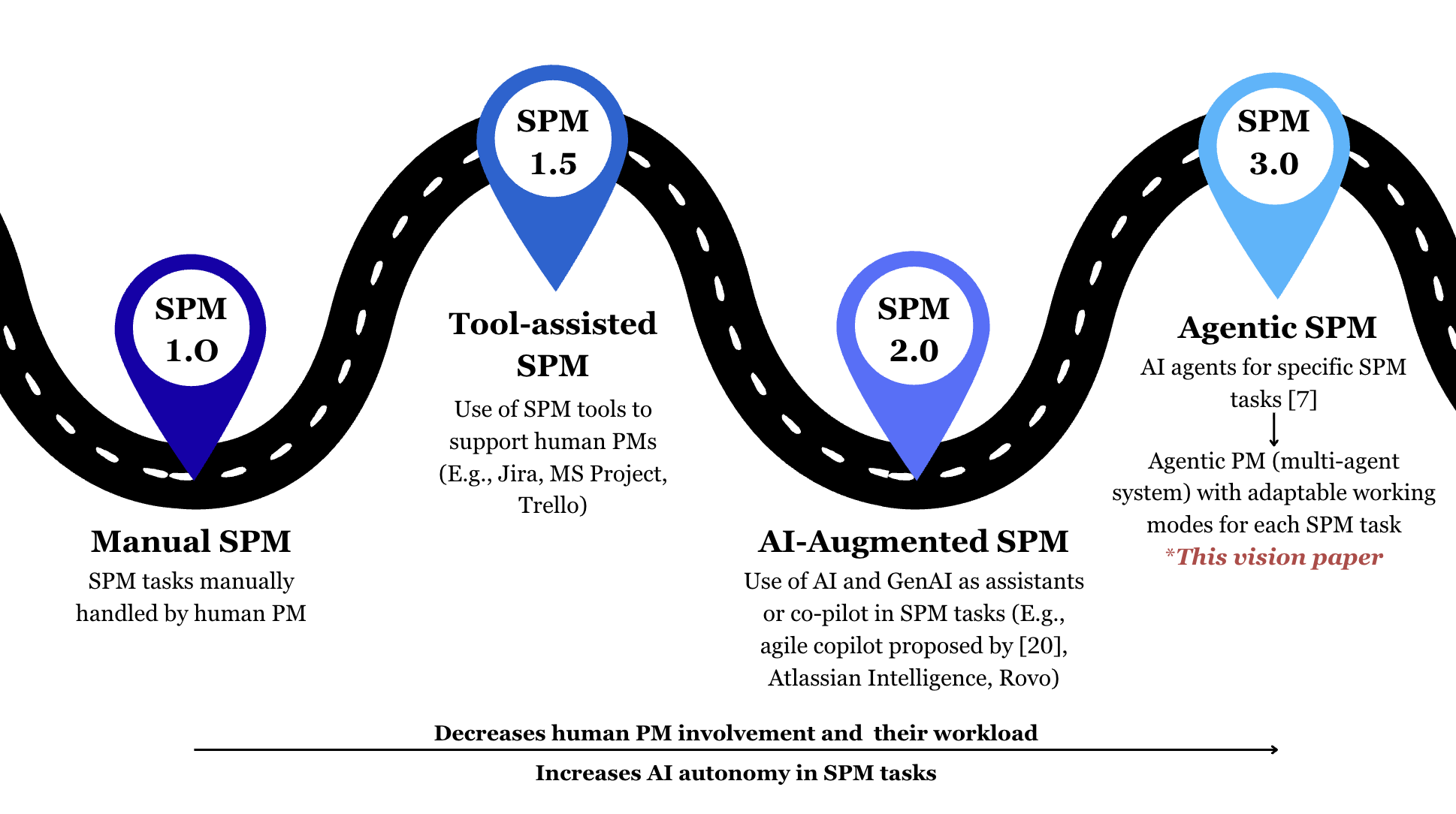}}
\caption{SPM with AI Roadmap (following \cite{Hassan2025})}
\Description{Illustration of SPM evolution with AI over the time}
\label{roadmap}
\end{figure}

\tcbset{
  researchbox/.style={
    colback=gray!5, colframe=black!60, arc=2mm,
    boxrule=0.4pt, left=6pt, right=6pt, top=4pt, bottom=4pt,
    before skip=8pt, after skip=8pt
  }
}
\begin{tcolorbox}[researchbox, title=\textbf{How SPM will evolve with AI to facilitate agentic SE?}]
\textbf{SPM with AI Roadmap} demonstrates the evolution of SPM with AI and the possibility for an \textit{\textbf{"Agentic SPM era (SPM 3.0)"}} having agentic PMs working under human control (Figure \ref{roadmap}). We cannot still predict a fully autonomous SPM future given the human-centric nature of SPM and few studies focusing on SPM evolution with LLMs.
\end{tcolorbox}
 
SPM transitioned from (manual) human-led SPM era (SPM 1.0) that used whiteboards, pens and papers to tool-supported SPM era (SPM 1.5), where SPM tools such as Microsoft Project, cloud-based platforms like Jira, and Trello \cite{Assalaarachchi2025} came to facilitate SPM activities, even supporting agile practices \cite{Micha2025}. With emerging AI capabilities, we then entered the AI-augmented SPM era (SPM 2.0), where AI became an assistant, or a co-pilot \cite{Dam2019, Hoda2023, Assalaarachchi2025a} in SPM. SPM tools started integrating AI (e.g., Rovo\footnote{https://www.atlassian.com/software/rovo}) to facilitate successful SPM \cite{Assalaarachchi2025a}. This evolution to SPM 2.0 is parallel to the AI-augmented SE (SE 2.0) era in the SE evolution proposed by \cite{Hassan2025}.

In our recent review study, we concluded with a vision about AI agents for SPM tasks and the possibility for an \textbf{\textit{agentic SPM}} era (SPM 3.0) \cite{Assalaarachchi2025a}. Now, we extend our vision aligning with Agentic SE (SE 3.0) transformations in \cite{Hassan2025} and propose \textit{\textbf{"agentic PM"}} as a multi-agent system to support the upcoming agentic SE era. Agentic PM can facilitate complex goals by perceiving the scenarios based on multiple data sources, make decisions, and manage tasks proactively once allowed to them by the human PM \cite{Acharya2025}. After executing the tasks, they can learn and adapt from the feedback of human PM and team members (Figure \ref{agenticSPM}). 

Along this evolution, we notice an increase in AI autonomy while automating more SPM tasks and positioning the human PM as a \textit{"coach"} who guides, supervises, and facilitates the team, including agentic teammates, towards successful project completion. Although \cite{Hassan2025} envision an autonomous SE (SE 4.0 and 5.0) era, we do not envision a fully autonomous SPM future or replacement of the human PM role now or in the near future. The human-centric nature of the PM role \cite{Abrahao2025}, strategic leadership, emotional intelligence, empathy, and ethical oversight crucial in SPM, still require the presence of a human PM. A recent article presents SPM as one of the outer loop activities in SDLC is expected to continue with human control assisted by AI rather than full automation \cite{Abrahao2025}. The recent \textit{future of jobs} report by the World Economic Forum also presents human-centric skills such as analytical thinking, creative thinking, agility, empathy, and active listening as core skills required in the future despite the AI evolutions \cite{WorldEconomicForum2025}. Furthermore, software management has been highlighted as one of the most under-researched activities in studies using Large Language Models (LLMs) for SE activities \cite{hou2024SLR}, suggesting much progress is needed before higher levels of AI autonomy can be expected in SPM. Therefore, we present our vision of an agentic PM as an assistant to human PM for SPM 3.0 era (in Section ~\ref{agenticPM}) with adaptable working modes and insights on human PM role evolution (in Section ~\ref{humanPMRole}).

\subsection{Agentic PM for SPM 3.0} \label{agenticPM}

\begin{figure} [th]
\centerline{\includegraphics[width=18pc]{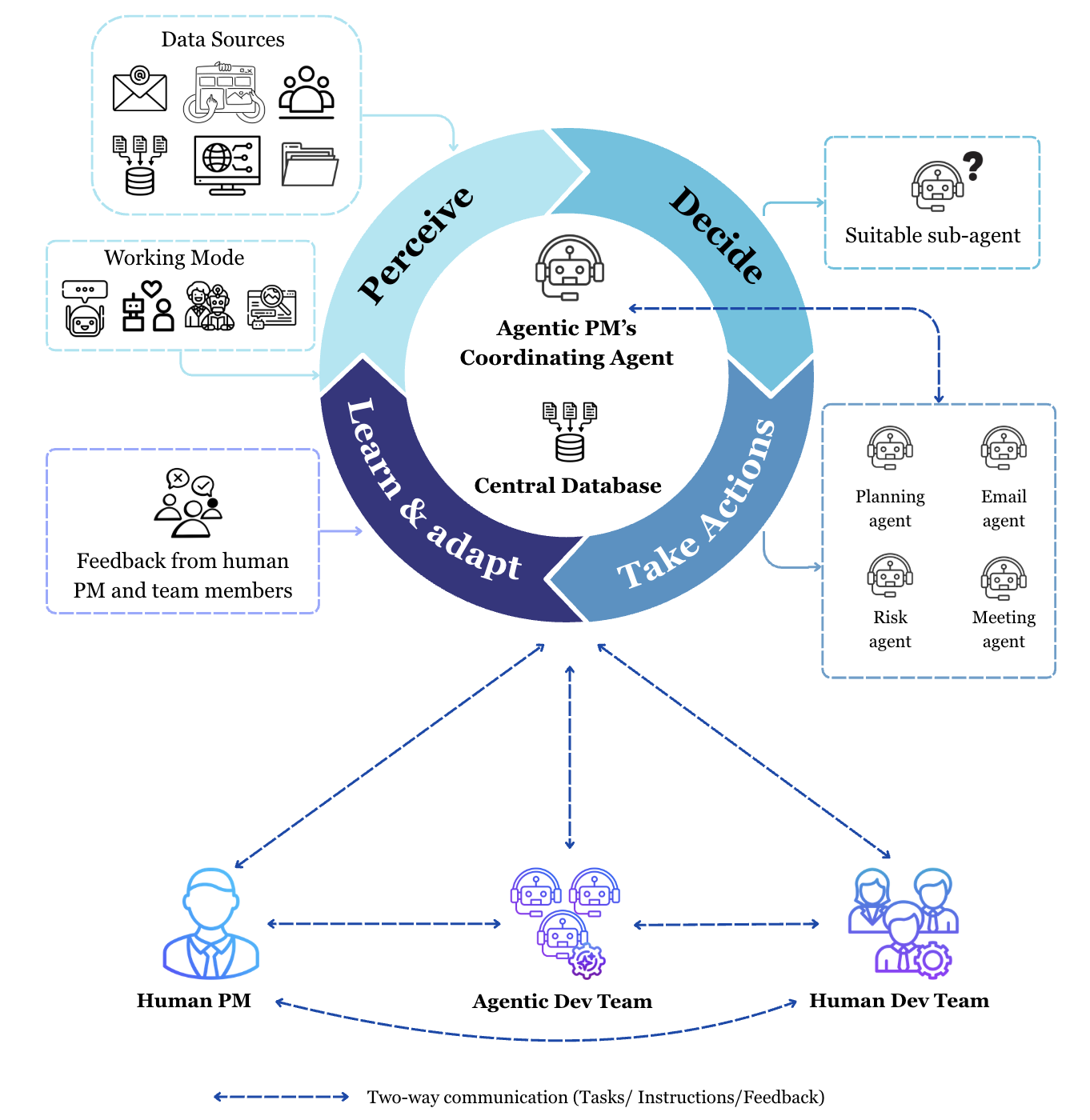}}
\caption{ Agentic PM as an assistant to human PM}
\Description{Diagram of how an agentic PM works as an assistant to human PM}
\label{agenticSPM}
\end{figure}

\begin{tcolorbox}[researchbox, title=\textbf{How should an ethical multi-agent project management assistant be designed and used while balancing autonomy?}]
\textbf{Agentic PM} is an ethical, multi-agent PM assistant that can \textit{perceive} the tasks at hand using multiple data sources, \textit{make decisions}, and \textit{take actions} based on the assigned working mode (Figure \ref{agenticSPM}). We propose four \textit{\textbf{working modes}} \textit{(guided AI-autonomy mode, supervised-AI mode, Human-AI collaborative mode, and AI-assisted mode)} to determine the autonomy level of the agentic PM based on task complexity and risk level. (Figure \ref{workingModes}). Agentic PM can also \textit{learn and adapt} from human feedback. 
\end{tcolorbox}

Our proposed agentic PM (multi-agent system) is expected to act as a \textit{“junior PM”} or an \textit{“intern PM”} on multiple SPM tasks collaboratively with the human PM and team members (human and agentic) represented in Figure \ref{agenticSPM}. 

The agentic PM will be composed of multiple sub-agents working on specialized tasks and a coordinating agent that centrally coordinate all sub-agents, and communications between sub-agents and with humans. The agentic PM will \textit{perceive} the tasks at hand using multiple data sources related to the scenario, and the assigned working mode. Then the coordinating agent can \textit{decide} by analyzing data and assign the tasks with the working mode to specialized sub-agent(s) to \textit{take actions} based on the assigned working mode. Once sub-agent executed the task, outcomes will be returned back to the coordinating agent. Human PM and the team members can then review those outcomes, approve/reject them, and provide feedback for the agentic PM to \textit{learn and adapt}. Human PM can also terminate the actions at anytime if needed via the coordinating agent. The coordinating agent will then communicate the decision and the feedback with the particular sub-agent to execute that action and learn for future actions. The agentic PM will include a central database to store all the outcomes, feedback, decisions, and related logs for governance and learning purposes.  

The four main capabilities of the agentic PM are described below.

\subsubsection{\textbf{Perceive}}
Agentic PM will first understand the assigned task and working mode by gathering information from multiple input sources such as SPM tools (e.g., Jira, Microsoft Project, Trello), documents (e.g., project plans, reports, software requirements specification), emails, user activities, databases, other agentic teammates, and online resources related to the scenario. However, we expect human PM to review and adhere to relevant national, organizational, or stakeholders' guidelines on data sharing and privacy when connecting the agentic PM with such data sources. This could minimize the ethical and data privacy concerns \cite{Assalaarachchi2025a}.

\begin{figure}
        \centerline{\includegraphics[width=20pc]{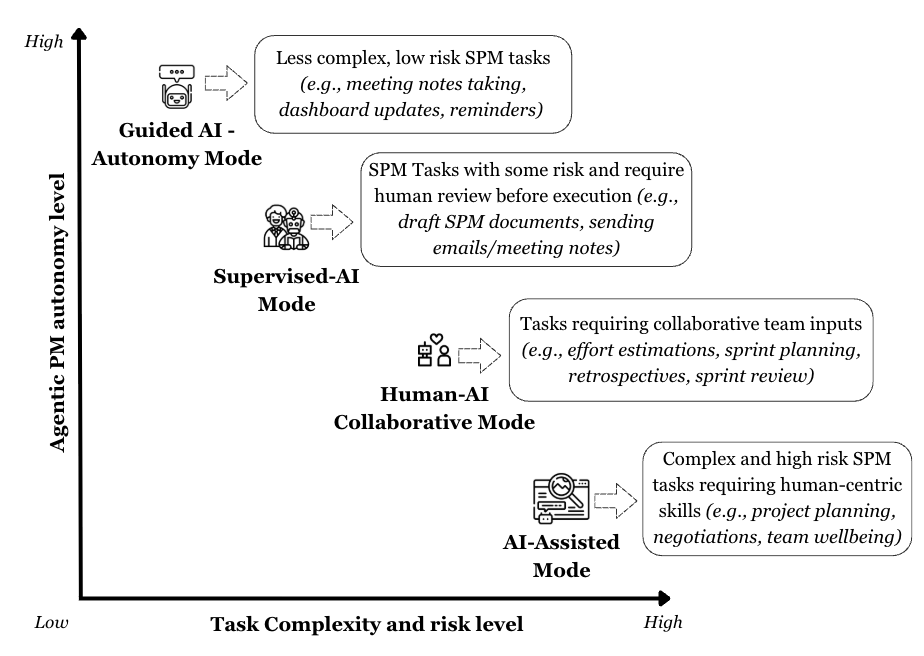}}
        \caption{Sample illustration of SPM task mapping to working modes in the agentic PM based on task complexity and risk}
        \Description{Sample illustration of SPM task mapping to working modes in the agentic PM based on task complexity and risk}
        \label{workingModes}
    \end{figure}

We adapt the concept of varying autonomy levels for AI agents proposed in \cite{Feng2025} to our context of the agentic PM and propose \textbf{\textit{four working modes}}, providing an opportunity for human PM to change the autonomy of the agentic PM considering the task complexity and risk level (Figure \ref{workingModes}). Complexity or difficulty of a task refers to the capability of human or agent to meet the specific requirements needed to perform a task (e.g., creativity, expertise knowledge, social skills, effort). Risk can be identified as the uncertainty of the outcomes associated with delegating the task and their impact which create the need for accountability towards the outcome. This will minimize the risks associated with giving full autonomy to an agentic teammate \cite{Feng2025, SPIEGLER2026103109} and align with the vision of keeping human control as suggested in \cite{hoda2025agenticsoftwareengineeringcode}. This would be similar to how we select different modes in LLMs (e.g., deep research mode in OpenAI for writing report) based on the nature of the task.  
   
    \begin{enumerate} [leftmargin=12pt]
        \item In the \textbf{guided AI-autonomy mode} (\includegraphics[height=1em]{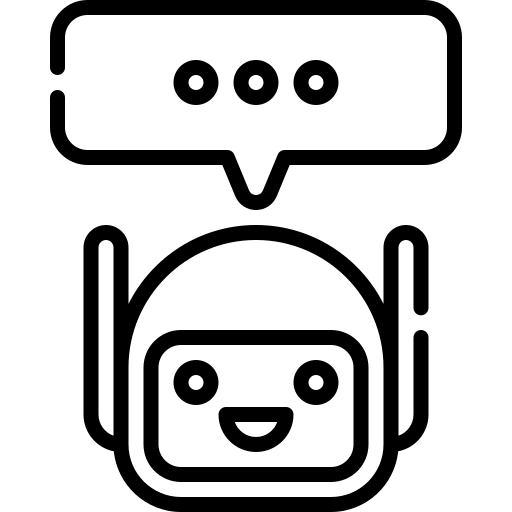}), the agentic PM can perform tasks with a higher autonomy but require human PM review upon final outcomes. We suggest less complex, and low risk tasks for this mode (e.g., meeting note taking, reminders). Human PM is accountable for carefully assessing the potential risks before assigning this working mode. This is somewhat similar to the idea of \textit{"Level 4: User as an approver"} proposed by \cite{Feng2025}. The human PM must provide detail instructions first and review the outcomes to approve or reject and provide feedback for the agentic PM to learn and adapt. Currently, we have AI meeting assistants such as Otter.ai, Fireflies, and AI integrations in video conferencing platforms such as Zoom and Microsoft Teams for meeting note taking. Although these AI meeting assistants can enhance the efficiency and productivity of PMs, possibility for concerns about data accuracy, misinterpretation, and ethical concerns such as data privacy have been reported \cite{Rebelo2025, SimpsonGrierson2025, Taylormoore}. Given the possibility for similar concerns in the agentic PM, we propose a guided AI-autonomy mode rather than a full autonomy level suggested in \textit{"Level 5: User as an observer"} of \cite{Feng2025}. The human PM oversight is encouraged for all outcomes, rather than over relying on the agentic PM \cite{SimpsonGrierson2025, Taylormoore}.
        
        \item In the \textbf{supervised-AI mode} (\includegraphics[height=1em]{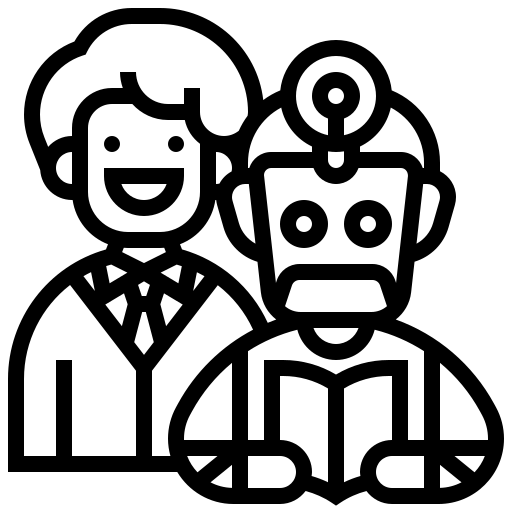}), we expect the human PM to provide the overview and expectations of the tasks to the agentic PM, and then the agentic PM could make drafts or prepare for the task (e.g., draft emails, reports). The agentic PM will consult the human PM time-to-time for feedback and fully execute the task once the human PM approve the modifications (e.g., sending finalized email or reports to stakeholders). This mode is proposed with reference to the \textit{"Level 3: User as a Consultant"} autonomy level suggested by \cite{Feng2025}. We can consider the deep research mode in OpenAI and Google Gemini as some related current examples \cite{Feng2025} that still need human oversight to review and modify those drafts to avoid hallucinations, include any organizational or project-specific information missed in the drafts, and to avoid data privacy concerns \cite{Assalaarachchi2025a}.
        
        \item \textbf{Human-AI Collaborative mode} (\includegraphics[height=1em]{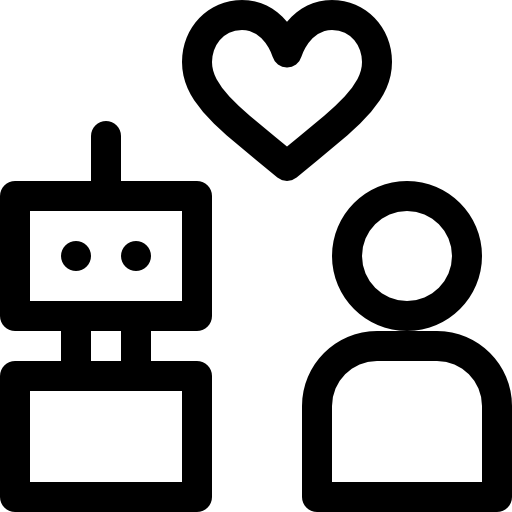}) will be suitable for tasks that require collaborative inputs from the whole team such as sprint planning, effort estimation, retrospectives, etc. Human PM and team members (both human and agentic) should interact with the agentic PM only as another collaborator supporting with data analysis and insights, while avoiding over reliance \cite{Assalaarachchi2025a}. Preliminary work in this area is already being done for agile effort estimation \cite{Bui2025}. This working mode aligns with the autonomy \textit{"Level 2: User as a collaborator"} proposed by \cite{Feng2025}.
        
        \item We suggest using \textbf{AI-Assisted mode} (\includegraphics[height=1em]{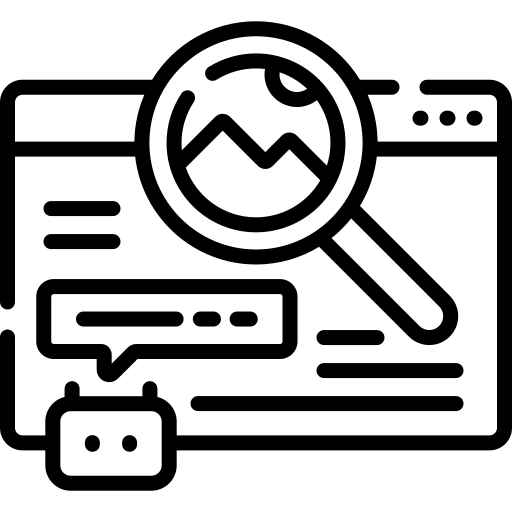}) when working with complex and strategic tasks such as project planning (wider impact upon whole project), negotiating, and mentoring which require more critical thinking and human-centric skills. Human PM needs to lead all activities under this mode, and the agentic PM works only as an assistant to data analysis and provide suggestions once requested by human PM. Through a global survey among PMs across various sectors, a recent PMI report also presents stakeholder management, project communication and project budgeting as the top three areas that have the lowest AI impact \cite{Nilsson2025}. This working mode is suggested based on the \textit{"Level 1: User as an operator"} autonomy level proposed by \cite{Feng2025}. Rovo, Microsoft Copilot, ChatGPT, Gemini are some current examples that works in a manner, which provide outcomes when users prompted them to. Still human PM needs to carefully review the agentic PM's outcomes to avoid hallucinations and analyses practical applicability of those suggestions.
    \end{enumerate}
    
    Table \ref{tab1} presents some examples of how an agentic PM is expected to work in each working mode\footnote{Note that some SPM tasks could be mapped to more than one working mode depending on the scenario (e.g., We state project planning as a suitable task under AI-assisted mode considering the high complexity, high risk, and need for more critical thinking required in most scenarios. But in some cases (e.g., small project with simple requirements), could assign Human-AI collaborative mode to the agentic PM in project planning). We present a preliminary vision to guide human PMs on ethical human-agent interactions with example scenarios which human PMs can adjust suited to the particular scenario.}, their success condition(s), anticipated risks requiring human PM oversight, existing examples, and their reported ethical incidents. 

\renewcommand{\arraystretch}{2}
\begin{table*} [t]
\caption{Examples for Agentic PM tasks performed in each working mode, their success condition(s), anticipated risks requiring human PM oversight, Existing Examples, and their reported ethical incidents}
\label{table}
\footnotesize
\begin{tabular*}{\textwidth}{@{\extracolsep{\fill}}
    >{\centering\arraybackslash}m{0.08\textwidth}
    >{\centering\arraybackslash}m{0.08\textwidth}
    >{\centering\arraybackslash}m{0.16\textwidth}
    >{\centering\arraybackslash}m{0.23\textwidth}
    >{\centering\arraybackslash}m{0.08\textwidth}
    >{\centering\arraybackslash}m{0.25\textwidth}
@{}}
\toprule
\textbf{Working Mode}& 
\textbf{Agentic PM task}& 
\textbf{Success condition(s)}&
\textbf{Anticipated risks (requiring Human PM oversight)}&
\textbf{Existing example(s)}&
\textbf{Ethical incident associated with existing examples}
\\ 
\midrule
Guided AI-Autonomy Mode (\includegraphics[height=1em]{source/autonomous.png})&
Meeting notes taking& 
Autonomously initiates meeting notes taking and shares with all attendees post meeting&
\begin{itemize} [leftmargin=12pt]
    \item Notes could include inaccuracies, wrong inferences, private, confidential or sensitive information that was not meant to be shared in written form
    \item Sharing meeting note with wrong stakeholders or missing some attendees when sharing
\end{itemize}&
Otter.ai, Fireflies, Granola, etc. \cite{Rebelo2025}&
Leakage of information from a confidential  company discussion with unintended parties by Otter.ai after a Zoom meeting, leading to cancellation of business deals \cite{AIIncidentDatabase2024}
\\ 
Supervised-AI Mode (\includegraphics[height=1em]{source/supervised.png})& 
SPM document creation& 
Create SPM documents for presentations and sharing with stakeholders&
\begin{itemize} [leftmargin=12pt]
    \item Missing important information within the document
    \item Documents might include inaccuracies, wrong inferences, private, confidential or sensitive information that was not meant to be shared in written form
\end{itemize}&
Deep research mode in OpenAI, Google Gemini, Perplexity etc. \cite{Feng2025} &
Two US judges have admitted the use of ChatGPT and Perplexity in drafting court filings that included fabricated court cases. Those court filings were subjected to judicial inquiry and retracted due to erroneous data \cite{AIIncidentDatabase2025}, Deloitte refunding \$440,000 worth consultancy fee to Australian government due to errors, and fabricated references in the report generated using Azure OpenAI GPT-4o \cite{NDTVWorld2025}
\\ 
Human-AI Collaborative Mode (\includegraphics[height=1em]{source/collaborative.png})& 
Sprint Planning& 
Analyses user stories, past project data and suggest possible user stories for the next sprint and story point estimations&
\begin{itemize} [leftmargin=12pt]
    \item Privacy concerns when sharing confidential project data
    \item Inaccuracies in suggestions when ignored human aspects such as team dynamics, skills, and workload
\end{itemize}&
Rovo \cite{Atlassian2025}&
Bui et al. \cite{Bui2025} introduce Software Effort Estimation Agent (SEEAgent) that provide justifications for the estimations to avoid hallucinations in general LLMs and consider human values crucial for tasks like effort estimation and sprint planning (also suggested in \cite{Assalaarachchi2025a, Hoda2023} 
\\ 
AI-Assisted Mode (\includegraphics[height=1em]{source/asisstant.png})& 
Project Planning& 
Analyse data related to project and provide insights in project planning&
\begin{itemize} [leftmargin=12pt]
    \item Privacy concerns when sharing project-specific or client-related details
    \item Generate some inaccurate or unrealistic suggestions
\end{itemize}&
Rovo \cite{Atlassian2025}, GenAI tools like Microsoft Copilot, ChatGPT etc.&
Google AI reported to produce inaccurate or harmful suggestions (e.g., presenting Barack Obama as the first Muslim U.S. President, glue as an ingredient for pizza) \cite{AIIncidentDatabase2024_a}
\\
\bottomrule
\end{tabular*}
\label{tab1}
\end{table*} 
    
Human PM can assign the suitable working mode by carefully assessing task complexity and risk associated with the assigned autonomy. In our preliminary vision, we consider these two common task-specific factors to provide a sample illustration to guide human PMs on mapping SPM tasks to suitable working mode (Figure~\ref{workingModes}). Human PM involvement should increase with the task complexity and risk level while reducing the autonomy of the agentic PM as proposed in \cite{Feng2025}. However, accountability for outcomes in all working modes should remain with the human PM as the ultimate decision maker and an ethical leader. This also supports the \textit{"whole of process"} vision and the \textit{principle of ethical alignment} for Agentic SE introduced in \cite{hoda2025agenticsoftwareengineeringcode}, in which all actions should be designed with ethics in mind and ultimately remain under human control. Also, our proposed agentic PM vision with the above four working modes supports the idea that activities such as SPM would stay under human control, assisted by AI, rather than full automation \cite{Abrahao2025, Lubars2019}. Therefore, we introduce only four working modes by adapting the autonomy level 1 to level 4 and excluding a fully autonomous mode like level 5 in \cite{Feng2025} for our agentic PM vision.  

\subsubsection{\textbf{Decide}}
The coordinating agent can analyse the task assigned, working mode, input sources, and similar past assignments to decide on the suitable task-specialized sub-agent(s) to assign the task. It will then communicate the task, working mode, and other information with that specific sub-agent to take actions.    

\subsubsection{\textbf{Take Actions}}    
Task-specialized sub-agents can then take actions allocated to them by the coordinating agent. Agents needs to consider the working mode since they will not take action alone by themselves in most working modes. They require approval or review of the human PM and team members in most modes. We suggest multiple interconnected task-specialized agents within the agentic PM system to perform several tasks based on the working mode assigned by the human PM \textit{(e.g., planning agent, email agent, meeting note agent, risk analysis agent, tasks management agent, etc.)}. These sub-agents can communicate with each other via the coordinating agent and the coordinating agent can escalate any issues or conflicts among agents to human PM for their review and decision. 
    
\subsubsection{\textbf{Learn and adapt}}   
Humans will have the capability to provide feedback to the agentic PM so it can learn and adapt in the next iterations. The agentic PM will have a central database to maintain logs on outcomes and feedback for future applications using reinforcement learning capability. This will be similar to how a human intern learns from PM's supervision and team members’ feedback. Humans can also provide feedback on outcomes' ethical compliance, enabling agentic PM to learn and minimize such issues in the next iterations.

Our proposed vision of an agentic PM with adaptable working modes can guide AI developers to create ethical agentic PMs and their users on responsible human-AI collaborations. It will also help in addressing the concerns such as clear task delegation, accountability, trust, and fear of job security that leads  human PMs to accept and adopt agentic PMs to work together with them.

\subsection{Human PM role evolution with Agentic SPM} \label{humanPMRole}

\begin{tcolorbox}[researchbox, title=\textbf{How should the human PM role evolve to facilitate responsible human-agent interactions?}]
Human PMs must upskill themselves and evolve their role into an \textit{\textbf{"ethical and strategic leader"}}, or a \textit{\textbf{“coach”}} who guides, supervises, and facilitates the team, including agentic teammates, towards successful project completion.
\end{tcolorbox}

Emerging literature suggests the need to redefine SE roles with the evolution of agentic AI in SE \cite{Hassan2025, hoda2025agenticsoftwareengineeringcode, Roychoudhury2025}. Similarly, we envision changes to the traditional role of the software PM when having an agentic PM as an assistant. In our proposed vision of agentic PM, we rely on the idea that human-AI collaboration is crucial for successful project completion while keeping all actions under human control \cite{hoda2025agenticsoftwareengineeringcode}. Hence, the human PM role will not be replaced, but rather augmented and assisted by agentic PMs as suggested earlier \cite{Hoda2023}. We propose an agentic PM as a junior PM or an intern PM working together with the development team (human and agentic teammates) and the human PM rather. This requires the human PM to facilitate not just human team members but also agentic PMs. Cognitive tasks such as \textit{project planning, strategic decision making, stakeholder management, and negotiations} rely on the human PM, who can be assisted by the agentic PM’s data analysis and insights \cite{Gress2025, Nilsson2025}. To align with the principles of \textit{ethics-by-design} and \textit{human, agentic, and socio-technical aspects} \cite{hoda2025agenticsoftwareengineeringcode}, the human PM will be responsible for maintaining \textit{ethical oversight} and improving \textit{human-agent collaborations} to get the full potential of agentic PM responsibly. 
    
With this expected shift, human PMs will need to master essential skills for responsible human-agent collaboration and strategic leadership. A recent practitioner discussion present the skills gap as a challenge that emerges agentic AI evolution \cite{Gress2025}. Through a review of similar literature \cite{Assalaarachchi2025a, White2024, hoda2025agenticsoftwareengineeringcode, Roychoudhury2025} and practitioner insights \cite{Gress2025, Nilsson2025, WorldEconomicForum2025}, we recommend human PMs enhancing skills such as:
\begin{itemize} [leftmargin=12pt]
    \item \textbf{Data interpretation -} Review and interpret outcomes from agentic PM, who works mostly on data analysis and suggestions.
    \item \textbf{Ethical oversight -} Ethics and data privacy knowledge to review agentic PM's outcomes and continuously monitor the agentic PM for ethical compliance \cite{Boinodiris2025}. Human PMs must adhere to  national, organizational and stakeholder ethical guidelines \cite{Assalaarachchi2025a}. 
    \item \textbf{Critical thinking -} The human PM acts as the ultimate decision maker for most non-trivial SPM tasks and requires critical thinking to make strategic decisions. Human PM should also critically evaluate the risks of different SPM tasks before assigning the suitable working mode for the agentic PM.
    \item \textbf{Interpersonal skills -} Skills such as emotional intelligence \cite{9769966}, empathy \cite{Gunatilake2025}, strategic leadership, and communication remain crucial for human PM in stakeholder management, negotiations, and conflict resolutions.
\end{itemize}

Human PMs must upskill themselves and evolve their role into a \textit{“coach”} who guides, supervises, and facilitates the team, including agentic teammates, towards successful project completion.

\section{Limitations and Future Work}
We acknowledge the limitation of this study as a preliminary vision without empirical evidences. We present this vision paper as a foundation for research on agentic AI and SPM to facilitate emerging agentic SE visions. Given the emerging nature of literature on agentic AI in SE and SPM domains, where most literature are yet pre-prints or vision papers, we plan to validate our vision empirically and fill the research gaps. We also acknowledge that our preliminary vision is limited to two task-specific factors (complexity and risk) when introducing a sample illustration of SPM task mapping to suitable working mode.  

As a next step, we will conduct a survey to understand the practitioners' perceptions towards the agentic SPM vision and the agentic PM with adaptable working modes. It will allow us to map the working modes to suitable SPM tasks and identify the factors influencing the task delegation decision to the agentic PM through empirical evidences. Based on the findings from the survey and follow-up interviews, we will develop a prototype of this agentic PM which we can later experiment with a software development team for validation. We expect to provide guidelines on responsible human-agent interactions based on those experiments. We also plan to explore practitioners' perceptions on expected human PM role changes and upskilling requirements with this SPM evolution. 

\section{A Research Agenda} \label{researchAgenda}

In addition to our empirical studies exploring parts of this preliminary vision, we present the following as future research opportunities for researchers in agentic SE and agentic SPM domains.
\begin{itemize} [leftmargin=12pt]
    \item \textbf{Factors affecting the acceptance of agentic PMs} - It is recommended to explore factors that support the adoption of agentic PMs to promote them to achieve project success. It is also crucial to understand the barriers for adoption and strategies to overcome them. This can lead to the successful implementation of agentic PMs while avoiding negative returns on investments \cite{Gartner2025}. It is also encouraged to explore organizational strategies to create awareness about agentic AI among employees, develop guidelines on responsible human-agent interactions, and facilitate necessary ethical or data security mechanisms. 
        
    \item \textbf{Software team members’ perceptions towards having an agentic PM} - Future research can explore team members' perspectives and experiences about agentic PMs in their team and how team members would collaborate with agentic PMs. It can also be expanded to explore collaborations between the agentic PM and other agentic teammates (multi-agentic SE teams) given the emerging research on agentic SE teammates \cite{Hassan2025, Li2025}.
        
    \item \textbf{Exploring the applicability of agentic PMs in agile methodologies} - Since our preliminary vision is not based on a particular SPM methodology, future researchers can explore the applicability of agentic PMs in popular agile methodologies such as Scrum \cite{Digital.ai2024}, using case studies with scrum teams. Since our proposed agentic PM vision also support agile values such as learning and adaptability, we recommend future empirical studies to validate or build upon our vision to have agentic PMs in agile team. 
        
    \item \textbf{Guidelines on responsible agentic PM development and usage} - With the evolution of agentic AI in SE, it is crucial to develop responsible agentic AI design and usage guidelines \cite{Boinodiris2025, hoda2025agenticsoftwareengineeringcode}. We provide the foundation by introducing various autonomy modes for different tasks, and responsible human-agent interactions in our agentic PM vision, which future researchers can build upon and develop guidelines to support working with agentic teammates. We also present the need for human oversight to ensure ethical compliance when working with an agentic PM, which can be further validated and explored in future studies with practitioner experiences.
        
    \item \textbf{Impact of agentic PMs on the early career PM roles} - We note the need for exploring the impact of agentic PMs on early career PM roles (interns/junior PMs) since the agentic PM is supposed to act in a similar role within the software team. Therefore, it is crucial to identify how early career PM roles should be redefined with new responsibilities and skills.

    \item \textbf{SPM education restructuring} - SPM education needs to be restructured, including agentic AI in curricula and evaluations based on skills required for the agentic SPM era. Aligning with the idea of building awareness about agentic AI through education, which is presented in \cite{hoda2025agenticsoftwareengineeringcode}, future research can be carried out on restructuring SPM education involving both practitioners and academics.
\end{itemize}

\section{Conclusion}
Emerging literature envision agentic SE (SE 3.0) era \cite{Hassan2025}, creating a need for transforming aspects of SDLC, like SPM, to facilitate agentic SE era \cite{hoda2025agenticsoftwareengineeringcode}. Therefore in this vision paper, we build the foundation by introducing \textbf{\textit{"SPM with AI Roadmap"}} to present evolution of SPM into agentic SPM (SPM 3.0) era supporting the vision of \cite{Hassan2025}. We build upon our prior vision of having AI agents for SPM tasks \cite{Assalaarachchi2025a}, and propose an ethical \textbf{\textit{"agentic PM"}} along with four \textbf{\textit{"working modes"}} that vary the autonomy of agentic PM for SPM tasks. These working modes provide guidance for human PMs in determining responsible human-agent interactions and the need for ethical oversight over the agentic PM's outcomes. We cannot predict a full autonomous SPM future or a replacement of the human PM role given the human-centric nature and lack of empirical works about AI capabilities in SPM \cite{Abrahao2025, hou2024SLR}. Therefore, we envision human PM to evolve their role to become an \textit{ethical and strategic leader} or a \textit{coach} with this SPM evolution. We present upskilling requirements for human PMs to support that role evolution.

Our proposed vision will guide AI developers in developing a \textit{trustworthy} agentic PMs and current (human) PMs on responsible human-agent interactions, possible role evolution, and upskilling requirements needed for the agentic SPM era. Since our study is limited to a preliminary vision from emerging literature, we invite future researchers in agentic SPM and agentic SE domains to carry out more empirical studies and build upon the vision as suggest in the research agenda (in section \ref{researchAgenda}).

\bibliographystyle{ACM-Reference-Format}
\bibliography{source/AgenticSPM}

@misc{Boinodiris2025,
author = {Boinodiris, Phaedra and Parker, Jon},
booktitle = {IBM},
title = {{The evolving ethics and governance landscape of agentic AI}},
url = {https://www.ibm.com/think/insights/ethics-governance-agentic-ai},
urldate = {2025-11-05},
year = {2025}
}

@article{hou2024slr,
author = {Hou, Xinyi and Zhao, Yanjie and Liu, Yue and Yang, Zhou and Wang, Kailong and Li, Li and Luo, Xiapu and Lo, David and Grundy, John and Wang, Haoyu},
title = {Large Language Models for Software Engineering: A Systematic Literature Review},
year = {2024},
issue_date = {November 2024},
publisher = {Association for Computing Machinery},
address = {New York, NY, USA},
volume = {33},
number = {8},
issn = {1049-331X},
url = {https://doi.org/10.1145/3695988},
doi = {10.1145/3695988},
journal = {Transactions on Software Engineering and Methodology},
month = dec,
articleno = {220},
numpages = {79}
}

@misc{Mori2025,
author = {Mori, Brunella},
booktitle = {LinkedIn},
title = {{Agentic AI in Project Management: The Future of Smart Decision-Making}},
url = {https://www.linkedin.com/pulse/agentic-ai-project-management-future-smart-brunella-mori-zg8tc/},
urldate = {2025-09-25},
year = {2025}
}

@misc{White2024,
archivePrefix = {arXiv},
arxivId = {2408.01768},
author = {White, Jules},
booktitle = {arXiv},
eprint = {2408.01768},
month = {aug},
title = {{Building Living Software Systems with Generative \& Agentic AI}},
url = {https://arxiv.org/abs/2408.01768},
year = {2024}
}

@misc{Li2025,
archivePrefix = {arXiv},
arxivId = {2507.15003},
author = {Li, Hao and Zhang, Haoxiang and Hassan, Ahmed E.},
booktitle = {arXiv},
eprint = {2507.15003},
month = {jul},
title = {{The Rise of AI Teammates in Software Engineering (SE) 3.0: How Autonomous Coding Agents Are Reshaping Software Engineering}},
url = {http://arxiv.org/abs/2507.15003},
year = {2025}
}

@article{Abrahao2025,
author = {Abrah{\~{a}}o, Silvia and Grundy, John and Pezz{\`{e}}, Mauro and Storey, Margaret-Anne and Tamburri, Damian A.},
doi = {10.1145/3715111},
issn = {1049-331X},
journal = {ACM Transactions on Software Engineering and Methodology},
month = {jun},
number = {5},
pages = {1--46},
title = {{Software Engineering by and for Humans in an AI Era}},
url = {https://dl.acm.org/doi/10.1145/3715111},
volume = {34},
year = {2025}
}

@misc{Karatas2025,
author = {Karatas, Emre},
booktitle = {Medium},
title = {{How the Agentic Approach Will Revolutionize Software Engineering}},
url = {https://medium.com/@emrekaratas-ai/how-the-agentic-approach-will-revolutionize-software-engineering-aea3b46e76b7},
urldate = {01/10/2025},
year = {2025}
}

@misc{Hassan2025,
archivePrefix = {arXiv},
arxivId = {2509.06216},
author = {Hassan, Ahmed E. and Li, Hao and Lin, Dayi and Adams, Bram and Chen, Tse-Hsun and Kashiwa, Yutaro and Qiu, Dong},
booktitle = {arXiv},
eprint = {2509.06216},
month = {sep},
title = {{Agentic Software Engineering: Foundational Pillars and a Research Roadmap}},
url = {http://arxiv.org/abs/2509.06216},
year = {2025}
}

@misc{Nilsson2025,
author = {Nilsson, Marly},
booktitle = {Project Management Institute, Inc},
title = {{The AI in Project Management Global Report: 1 Year Later, 2025 and Beyond}},
url = {https://www.projectmanagement.com/articles/1049056/the-ai-in-project-management-global-report--1-year-later--2025-and-beyond#_},
urldate = {2025-10-01},
year = {2025}
}

@misc{Masood2025,
author = {Masood, Adnan},
booktitle = {Medium},
title = {{AI in Project Management — How Generative and Agentic AI Are Redefining Strategy, Execution, and Value Delivery}},
url = {https://medium.com/@adnanmasood/ai-in-project-management-how-generative-and-agentic-ai-are-redefining-strategy-execution-and-ccfd45229e7b},
urldate = {27/09/2025},
year = {2025}
}

@misc{Gartner2025,
author = {Gartner},
booktitle = {Gartner},
title = {Gartner Predicts Over 40\% of Agentic AI Projects Will Be Canceled by End of 2027},
url = {{https://www.gartner.com/en/newsroom/press-releases/2025-06-25-gartner-predicts-over-40-percent-of-agentic-ai-projects-will-be-canceled-by-end-of-2027}},
urldate = {03/10/2025},
year = {2025}
}

@inproceedings{Dam2019,
author = {Dam, Hoa Khanh and Tran, Truyen and Grundy, John and Ghose, Aditya and Kamei, Yasutaka},
title = {Towards effective AI-powered agile project management},
year = {2019},
publisher = {IEEE Press},
url = {https://doi.org/10.1109/ICSE-NIER.2019.00019},
doi = {10.1109/ICSE-NIER.2019.00019},
booktitle = {Proceedings of the 41st International Conference on Software Engineering: New Ideas and Emerging Results},
pages = {41–44},
numpages = {4},
address = {Piscataway, NJ, USA},
series = {ICSE-NIER '19}
}

@inproceedings{Kardum2025,
author = {Kardum, Benjamin and Car, {\v{Z}}eljka},
booktitle = {2025 MIPRO 48th ICT and Electronics Convention},
doi = {10.1109/MIPRO65660.2025.11132036},
isbn = {979-8-3315-3597-1},
month = {jun},
pages = {1862--1867},
publisher = {IEEE},
title = {{Exploring Stakeholders' Perspectives on the Impact of Generative AI on IT Project Management Roles}},
url = {https://ieeexplore.ieee.org/document/11132036/},
year = {2025}
}

@article{Hoda2023,
author = {Hoda, Rashina and Dam, Hoa and Tantithamthavorn, Chakkrit and Thongtanunam, Patanamon and Storey, Margaret-Anne},
doi = {10.1109/MS.2023.3268725},
issn = {0740-7459},
journal = {IEEE Software},
month = {jul},
number = {4},
pages = {106--109},
title = {{Augmented Agile: Human-Centered AI-Assisted Software Management}},
url = {https://ieeexplore.ieee.org/document/10176159/},
volume = {40},
year = {2023}
}

@ARTICLE{Assalaarachchi2025a,
  author={Assalaarachchi, Lakshana Iruni and Masood, Zainab and Hoda, Rashina and Grundy, John},
  journal={IEEE Software}, 
  title={Generative AI for Software Project Management: Insights from a Review of Software Practitioner Literature}, 
  year={2025},
  pages={1-8},
  note={(Early Access)},
  doi={10.1109/MS.2025.3619936}
}

@article{Hughes2025,
author = {Hughes, Laurie and Dwivedi, Yogesh K. and Li, Keyao and Appanderanda, Mandanna and Al-Bashrawi, Mousa Ahmad and Chae, Inyoung},
doi = {10.1080/1097198X.2025.2524286},
issn = {1097-198X},
journal = {Journal of Global Information Technology Management},
month = {jul},
number = {3},
pages = {175--185},
title = {{AI Agents and Agentic Systems: Redefining Global it Management}},
url = {https://www.tandfonline.com/doi/full/10.1080/1097198X.2025.2524286},
volume = {28},
year = {2025}
}

@misc{Gress2025,
author = {Gress, Brandes},
booktitle = {Wrike},
title = {{AI agents in project management: The new force behind high-performing teams}},
url = {https://www.wrike.com/blog/ai-agents-in-project-management/},
urldate = {13/10/2025},
year = {2025}
}

@misc{Feng2025,
archivePrefix = {arXiv},
arxivId = {2506.12469},
author = {Feng, K. J. Kevin and McDonald, David W. and Zhang, Amy X.},
booktitle = {arXiv},
eprint = {2506.12469},
month = {jul},
title = {{Levels of Autonomy for AI Agents}},
url = {http://arxiv.org/abs/2506.12469},
year = {2025}
}

@article{electronics14010087,
author = {Cinkusz, Konrad and Chudziak, Jaros{\l}aw A and Niewiadomska-Szynkiewicz, Ewa},
doi = {10.3390/electronics14010087},
issn = {2079-9292},
journal = {Electronics},
number = {1},
title = {{Cognitive Agents Powered by Large Language Models for Agile Software Project Management}},
url = {https://www.mdpi.com/2079-9292/14/1/87},
pages = {1-33},
volume = {14},
year = {2025}
}

@article{Sapkota2026,
author = {Sapkota, Ranjan and Roumeliotis, Konstantinos I. and Karkee, Manoj},
doi = {10.1016/j.inffus.2025.103599},
issn = {15662535},
journal = {Information Fusion},
month = {feb},
pages = {103599},
title = {{AI Agents vs. Agentic AI: A Conceptual taxonomy, applications and challenges}},
url = {https://linkinghub.elsevier.com/retrieve/pii/S1566253525006712},
volume = {126},
year = {2026}
}

@techreport{WorldEconomicForum2025,
author = {{World Economic Forum}},
institution = {{World Economic Forum}},
title = {{Future of Jobs Report}},
url = {https://www.weforum.org/reports/the-future-ofjobs-report-2025/},
year = {2025}
}

@article{Acharya2025,
author = {Acharya, Deepak Bhaskar and Kuppan, Karthigeyan and Divya, B.},
doi = {10.1109/ACCESS.2025.3532853},
issn = {2169-3536},
journal = {IEEE Access},
pages = {18912--18936},
title = {{Agentic AI: Autonomous Intelligence for Complex Goals—A Comprehensive Survey}},
url = {https://ieeexplore.ieee.org/document/10849561/},
volume = {13},
year = {2025}
}

@article{Assalaarachchi2025,
author = {Assalaarachchi, L. I. and Liyanage, M. P. P. and Hewagamage, C.},
doi = {10.12821/ijispm130204},
issn = {2182-7788},
journal = {International Journal of Information Systems and Project Management},
month = {apr},
number = {2},
pages = {1--20},
title = {{A framework of critical success factors of cloud based project management software adoption}},
url = {https://revistas.uminho.pt/index.php/ijispm/article/view/6449},
volume = {13},
year = {2025}
}

@misc{Micha2025,
author = {Micha{\l}},
booktitle = {FutureCode IT Consulting},
title = {{The Evolution of Project Management Tools: From Gantt Charts to AI-Powered Platforms}},
url = {https://future-code.dev/en/blog/the-evolution-of-project-management-tools/},
urldate = {22/10/2025},
year = {2025}
}

@misc{hoda2025agenticsoftwareengineeringcode,
      title={Toward Agentic Software Engineering Beyond Code: Framing Vision, Values, and Vocabulary}, 
      author={Rashina Hoda},
      year={2025},
      eprint={2510.19692},
      archivePrefix={arXiv},
      primaryClass={cs.SE},
      url={https://arxiv.org/abs/2510.19692},
      note = {(Accepted for AGENT Workshop at ICSE2026)},
}

@techreport{Digital.ai2024,
author = {Digital.ai},
institution = {Digital.ai},
title = {{17th State of Agile Report}},
url = {https://info.digital.ai/rs/981-LQX-968/images/RE-SA-17th-Annual-State-Of-Agile-Report.pdf},
year = {2024}
}

@misc{Bui2025,
archivePrefix = {arXiv},
arxivId = {2509.14483},
author = {Bui, Thanh-Long and Dam, Hoa Khanh and Hoda, Rashina},
eprint = {2509.14483},
month = {sep},
title = {{An LLM-based multi-agent framework for agile effort estimation}},
note = {(Presented at 40th IEEE/ACM International Conference on Automated Software Engineering)},
url = {http://arxiv.org/abs/2509.14483},
year = {2025},
}

@misc{Roychoudhury2025,
archivePrefix = {arXiv},
arxivId = {2502.13767},
author = {Roychoudhury, Abhik and Pasareanu, Corina and Pradel, Michael and Ray, Baishakhi},
eprint = {2502.13767},
month = {sep},
title = {{Agentic AI Software Engineers: Programming with Trust}},
url = {http://arxiv.org/abs/2502.13767},
year = {2025}
}

@ARTICLE{9769966,
  author={Madampe, Kashumi and Hoda, Rashina and Grundy, John},
  journal={IEEE Transactions on Software Engineering}, 
  title={The Emotional Roller Coaster of Responding to Requirements Changes in Software Engineering}, 
  year={2023},
  volume={49},
  number={3},
  pages={1171-1187},
  doi={10.1109/TSE.2022.3172925}
}

@article{Gunatilake2025,
author = {Gunatilake, Hashini and Grundy, John and Hoda, Rashina and Mueller, Ingo},
title = {The Role of Empathy in Software Engineering - A Socio-Technical Grounded Theory},
year = {2025},
publisher = {Association for Computing Machinery},
address = {New York, NY, USA},
issn = {1049-331X},
url = {https://doi.org/10.1145/3768315},
doi = {10.1145/3768315},
note = {Just Accepted},
journal = {ACM Trans. Softw. Eng. Methodol.},
month = sep,
note = {(Just Accepted)}
}

@article{SPIEGLER2026103109,
author = {Spiegler, Simone and Hoda, Rashina and Pant, Aastha},
doi = {https://doi.org/10.1016/j.techsoc.2025.103109},
issn = {0160-791X},
journal = {Technology in Society},
pages = {103109},
title = {{Images of AI: How AI practitioners view the impact of Artificial Intelligence on society, now and in the future}},
url = {https://www.sciencedirect.com/science/article/pii/S0160791X25002994},
volume = {84},
year = {2026}
}

@misc{Rebelo2025,
author = {Rebelo, Miguel},
booktitle = {Zapier},
title = {{The 9 best AI meeting assistants in 2025}},
url = {https://zapier.com/blog/best-ai-meeting-assistant/},
urldate = {06/11/2025},
year = {2025}
}

@misc{SimpsonGrierson2025,
author = {{Simpson Grierson}},
booktitle = {Simpson Grierson},
title = {{Directors take note: the legal pitfalls of AI in the boardroom}},
url = {https://www.simpsongrierson.com/insights-news/legal-updates/directors-take-note-the-legal-pitfalls-of-ai-in-the-boardroom},
year = {2025}
}

@misc{Taylormoore,
author = {Taylormoore, Melissa and {Cohen, Joel M.Davies}, Mark},
booktitle = {White & Case},
title = {{When every word is recorded: AI meeting tools and the new governance risks}},
url = {https://www.whitecase.com/insight-alert/when-every-word-recorded-ai-meeting-tools-and-new-governance-risks},
year = {2025},
urldate = {07/11/2025}
}

@misc{AIIncidentDatabase2024,
author = {{AI Incident Database}},
booktitle = {AI Incident Database},
title = {{Incident 811: AI-Powered Transcription Services Allegedly Leak Confidential Workplace Discussions}},
url = {https://incidentdatabase.ai/cite/811/},
urldate = {07/11/2025},
year = {2024}
}

@techreport{MIT2025,
author = {MIT NANDA},
institution = {MIT},
title = {{The GenAI Divide: State of AI in Business 2025}},
url = {https://mlq.ai/media/quarterly_decks/v0.1_State_of_AI_in_Business_2025_Report.pdf},
year = {2025}
}

@misc{Atlassian2025,
author = {Atlassian},
booktitle = {Atlassian},
title = {{Bring the power of human and AI collaboration to every team}},
url = {https://www.atlassian.com/software/rovo/features},
urldate = {07/11/2025},
year = {2025}
}

@misc{AIIncidentDatabase2025,
author = {{AI Incident Database}},
booktitle = {AI Incident Database},
title = {{Incident 1252: Judges in New Jersey and Mississippi Admit AI Tools Produced Erroneous Federal Court Filings}},
url = {https://incidentdatabase.ai/cite/1252/},
urldate = {07/11/2025},
year = {2025}
}

@misc{NDTVWorld2025,
author = {{NDTV World}},
booktitle = {NDTV World},
title = {{Deloitte's AI Fallout Explained: The \$440,000 Report That Backfired}},
url = {https://www.ndtv.com/world-news/deloittes-ai-fallout-explained-the-440-000-report-that-backfired-9417098},
urldate = {07/11/2025},
year = {2025}
}

@misc{AIIncidentDatabase2024_a,
author = {{AI Incident Database}},
booktitle = {AI Incident Database},
title = {{Incident 693: Google AI Reportedly Delivering Confidently Incorrect and Harmful Information}},
url = {https://incidentdatabase.ai/cite/693/},
urldate = {07/11/2025},
year = {2024}
}

@inproceedings{Lubars2019,
author = {Lubars, Brian and Tan, Chenhao},
title = {Ask not what AI can do, but what AI should do: towards a framework of task delegability},
booktitle = {Proceedings of the 33rd International Conference on Neural Information Processing Systems},
year = {2019},
publisher = {Curran Associates Inc.},
address = {Red Hook, NY, USA},
articleno = {6},
numpages = {11}
}

\end{document}